# The skyrmion switch: turning magnetic skyrmion bubbles on and off with an electric field

Marine Schott[1,2,3,4], Anne Bernand-Mantel[1,2]*, Laurent Ranno[1,2], Stéfania Pizzini[1,2], Jan Vogel[1,2], Hélène Béa[1,3,4], Claire Baraduc[1,3,4], Stéphane Auffret[1,3,4], Gilles Gaudin[1,3,4] and Dominique Givord[1,2]

[1] Univ. Grenoble Alpes, F-38000 Grenoble, France

[2] CNRS, Inst. NEEL, F-38000 Grenoble, France

[3] CNRS, SPINTEC, F-38000 Grenoble, France

[4] CEA, INAC-SPINTEC, F-38000 Grenoble, France

*anne.bernand-mantel@neel.cnrs.fr

**Nanoscale magnetic skyrmions are considered as potential information carriers for future spintronics memory and logic devices. Such applications will require the control of their local creation and annihilation, which involves so far solutions that are either energy consuming or difficult to integrate. Here we demonstrate the control of skyrmion bubbles nucleation and annihilation using electric field gating, an easily integrable and potentially energetically efficient solution. We present a detailed stability diagram of the skyrmion bubbles in a Pt/Co/oxide trilayer and show that their stability can be controlled via an applied electric field. An analytical bubble model, with the Dzyaloshinskii-Moriya interaction imbedded in the domain wall energy, account for the observed electrical skyrmion switching effect. This allows us to unveil the origin of the electrical control of skyrmions stability and to show that both magnetic dipolar interaction and the Dzyaloshinskii-Moriya interaction play an important role in the skyrmion bubble stabilization. Keywords : room temperature, magnetic switching, skyrmions, electric field, thin films, spintronics, perpendicular magnetic anisotropy**

Magnetic skyrmions are swirling spin textures with nanoscale dimensions and a non-trivial topology. They were studied theoretically more than two decades ago[1,2] and observed at low temperature in the form of hexagonal lattices in non-centrosymmetric crystals[3,4] and magnetic multilayers[5,6]. Recently several groups have reported the observation of magnetic skyrmions at room temperature (RT) in conventional transition-metal-based magnetic multilayers[7–14]. In some systems, the skyrmions present relatively large dimensions (~1 μm) and are called 'skyrmion bubbles'. While classical bubbles are stabilized by the magnetic dipolar energy, a skyrmion bubble is stabilized by both dipolar and Dzyaloshinskii-Moriya interaction (DMI) energies. A second difference between classical bubbles and skyrmionic bubbles is their homo-chirality which allows their current driven motion via spin-orbit torques[7,9,13,14]. This discovery has triggered a growing interest for the use of magnetic skyrmions as building blocks for memories[15–21] which is reminiscent of the extensive research on magnetic bubble memories in the 1970s[22]. Such applications will require creating, deleting and moving skyrmions. The recent observation of current induced displacement of skyrmion bubbles using relatively low current densities reported at RT[9,13,14,23] is very promising. The creation and annihilation of skyrmions have been addressed theoretically[17,24,25] and experimentally[6,26–28] by different techniques such as spin transfer torque[6,17,25], heat[24,28] and strain[27]. However, those techniques are either energy consuming or difficult to integrate in functional devices. In analogy to what has been developed to switch small ferromagnetic elements[29], the use of electric field gating to manipulate skyrmions offers several advantages: the low power consumption, the possibility to act locally and an easy integration. Furthermore, a major issue of current induced nucleation, namely, unwanted current displacement of skyrmions during the writing process, is naturally absent using electric field writing. Those advantages have motivated some theoretical works concerning the electric control of skyrmions[20,30–32]. The only experimental attempt for skyrmion electrical switching has been carried out in an epitaxial material and at low temperature[26]. In this letter, we report on the observation of thermally activated nucleation and annihilation of magnetic skyrmion bubbles at RT and we demonstrate the possibility to electrically switch them on and off under a constant biasing magnetic field.

We have selected for this study a $Pt/Co/oxide$ trilayer system where large DMI energy values have been reported[33–36]. In addition, we have chosen thicknesses for which the $Co$ layer is close to the ferromagnetic-paramagnetic transition at RT, as enhanced electric field effects have been observed in this conditions[37]. The $Co$ layer presents a slight thickness gradient (2%/mm) created by the oxidation of a wedge shaped $Al$ top layer (see Figure 1b). A high-k $HfO_2$ dielectric layer and a transparent top Indium Tin oxide ($ITO$) electrode have been deposited on top of the $Pt/Co/oxide$ trilayer, allowing magnetic characterization using magneto-optical Kerr effect under applied electric fields.

The magnetic domains are imaged at RT using a Kerr microscope in polar geometry. At low magnetic field a spontaneously demagnetized labyrinthine state is observed (Figure 1a). The formation of labyrinthine domains is the consequence of a balance between competing energies: the magnetic dipolar energy and the domain wall (DW) energy. A variation of the labyrinthine domain width $L$ is observed along the wedge direction (Figure 1c). To understand this variation we have used an analytical model[38,39] describing for ultrathin films the labyrinthine domain width $L = \alpha.t.\exp(\pi L_0/t)$ as function of the magnetic film thickness $t$, the characteristic dipolar length $L_0 = \sigma_W/\mu_0.M_s^2$, the saturation magnetization $M_S$, the DW energy $\sigma_W$ and $\alpha$ a numerical constant (see Methods). The experimental value of the DW energy $\sigma_W$ can be calculated from this equation with the measured values of $L$, t and $M_S$ (see supplementary S1-S3). At position $p_0$, $L = 2.9$ μm, $t = 0.47$ nm, $M_S = 0.93$ MA/m and we find $\sigma_W = 1.44$ mJ/m². This value can be compared to the classical Bloch wall energy $\sigma_w^{Bloch} = 4\sqrt{A_{ex}K_{eff}}$. Using the experimental effective anisotropy value $K_{eff} = 0.30$ MJ/m³ (see supplementary S4) and an exchange constant $A_{ex}$ in the $5 - 15$ pJ/m range we find $\sigma_w^{Bloch} = 5 - 8$ mJ/m² which is much larger than the experimental DW energy. The deviation from the Bloch formula can be explained by the presence of interfacial DMI in the trilayer which reduces $\sigma_W$ according to the expression[13,40]: $\sigma_W = 4\sqrt{A_{ex}K_{eff}} - \pi D$ where D is the DMI energy in mJ/m². We deduce a DMI term $D = 1 - 2$ mJ/m² which is consistent with the values obtained in similar $Pt/Co/oxide$ trilayers [33–36]. The low DW energy in our system is at the origin of the observed spontaneous demagnetization of the layer at RT on a few second timescale. This thermally activated demagnetization is due to a combination of spontaneous nucleation of magnetic domains and thermally activated motion of domain walls under zero applied magnetic field [(see supplementary S5)](). Such thermal movement of labyrinthine magnetic domains has been reported previously in ultrathin films where the DW energy was lowered by the proximity of the spin reorientation transition[41,42]. This DW mobility, in combination with the high nucleation rate, allows reaching an equilibrium demagnetized state in a few seconds in the major part of the image of Figure 1a (between $p_0$ and $p_3$). Above $p_0$, the demagnetized state starts to deviate from the equilibrium one and we decided not to extract the domain width in this region. We now discuss the origin of the decrease by a factor of 3 of the domain width $L$ along the wedge (Figure 1c). The $Co$ thickness $(t)$ variation is of the order of 1% along the considered region and its effect on $L$ in the expression $L = \alpha.t.exp(\frac{\pi\sigma_W}{\mu_0.M_s^2}/t)$ is negligible. However, as $t$ decreases, a reduction of $M_s$ by 8.4% is observed between $p_0$ and $p_3$ due to a variation of $T_C$ with $t$ and the proximity of $T_C$ at the observation temperature[43] (see supplementary S2). According to its analytical expression, the domain width $L$ should increase when $M_s$, and thus the magnetic dipolar energy, decreases. However, the opposite is observed in Figure 1c. We deduce that a strong decrease of the DW energy $\sigma_W$, over-compensating the $M_S$ variation is at the origin of the decrease of $L$. If we assume a constant DMI energy term, the variation of the DW energy $\sigma_W = 4\sqrt{A_{ex}K_{eff}} - \pi D$ must be related to $A_{ex}$ and $K_{eff}$ variations. This is confirmed by our experimental observations: a reduction of $K_{eff}$ with $t$ is measured (see supplementary S3) and a diminution of $A_{ex}$ is expected from the $M_S$ decrease with $t$[44,45].

When a perpendicular magnetic field larger than 0.1 mT is applied, micron-sized bubbles appear (Figure 1a). This transition from a labyrinthine pattern into a bubble lattice with applied magnetic field is unusual for a classical bubble system. In the classical case the magnetic domains would shrink in width and length but the number of magnetic objects would remain constant[38], leaving very few bubbles. This situation is observed on the right side of the image in Figure 1a. As we go from right to left, the increasing density of bubbles domains indicates the presence of thermally activated nucleation. This thermally activated process is promoted by the reduction of the DW energy $\sigma_w$, due to the presence of DMI. This implies a skyrmionic nature of the nucleated bubbles, which is confirmed by the observation of their unidirectional motion against electron flow, as expected for this system (see supplementary S6). As the nucleation of a skyrmion bubble by coherent rotation of a 1 μm domain is too high in energy ($KV > 10^4 k_B T_{293K}$) the mechanism for the observed skyrmion bubbles creation has to involve expansion from a nanoscale skyrmion. We can estimate the nucleation energy of a skyrmion bubble when its diameter is reduced to the DW width $\delta_w$ by $\sigma_w.t.2.\pi.\delta_w \sim 10 k_B T_{293K}$. Following a Néel Brown model, the RT nucleation rate of such skyrmion bubbles is high (~0.5 MHz for $\tau_0 = 0.1\ ns$). To estimate whether the nucleated skyrmion bubble is expected to annihilate or to grow and reach its equilibrium diameter, we have calculated analytically the energy of an isolated skyrmion bubble using a "thin wall" model which is valid for negligible DW width ($\delta_w \sim 10$ nm in our case) compared to the bubble diameter (few $\mu m$) (see Methods). The skyrmion bubble energy $E_{sb}(t, M_s, \sigma_w)$ is the sum of DW, Zeeman and dipolar terms. The energy of an individual bubble relative to the saturated state can be written as:

$$\Delta E_{sb}(R, t, M_s, \sigma_w) = \sigma_w.t.2.\pi.R + 2\mu_0 M_s.H.t.\pi.R^2 - \mu_0 M_s^2.\pi.t^3.I(d)$$

where $\sigma_w$ is the DW energy, $t$ the layer thickness, $R$ the bubble radius, $M_s$ the saturation magnetization, $\mu_0 H$ the applied magnetic field, and $I(d)$ the stray field energy gain where $d = 2R/t$ (see Methods). The skyrmion bubble energy $\Delta E_{sb}$ versus the bubble diameter (Figure 2.g), obtained using the experimental $t, M_s$ and $\sigma_w$ parameters at positon $p_1$, presents two local extrema. The existence of a local energy minimum in the model is consistent with the experimental observation of an equilibrium diameter for the skyrmion bubbles. To reach this equilibrium diameter, a nucleated nanoscale skyrmion will have to grow and overcome a nucleation energy barrier $E_n$, which corresponds to the local maximum. We define the annihilation energy barrier $E_a$ as the difference between the local maximum and minimum (see Figure 2.g), assuming that the lower energy path for annihilation is the compression[46–49]. The calculated nucleation and annihilation energies are shown in Figure 1d,e. For this calculation, we have used the $M_s$, $t$ and $\sigma_W$ values and their variations along the wedge estimated experimentally using a combination of magneto-optical Kerr and VSM-SQUID magnetometry (see supplementary S1-S3). As our characterization setup was adapted to study the 100 ms to few s timescale, we draw as a guide to the eye, a white line indicating the RT thermal activation frontier, $E/k_B T_{293K} = 23$ (value at witch nucleation events are occurring at a rate $\Gamma > 1$ Hz). The analytical skyrmion bubble model reproduces several features observed in Figure 1a. In Figure 1d, for a constant magnetic field, the increase of the nucleation barrier $E_n$ with the DW energy explains the decrease in the bubble density observed in Figure 1a as the $Co$ gets thicker. In the zone indicated by the number 2 in Figure 1d, skyrmion bubbles are observed despite the increase of $E_n$ above the thermal activation frontier. Those skyrmion bubbles are metastable skyrmions bubbles nucleated at lower magnetic field or stabilized by local magnetic inhomogeneities, which might locally lower the nucleation barrier. In the position indicated by the number 1 in Figure 1d,e, both nucleation and annihilation are expected with a rate $\Gamma > 1$ Hz. This thermal instability of the skyrmion bubbles is observed in a time dependent measurement (see supplementary S7). On the contrary, the skyrmion bubbles present a high thermal stability ($E_a/kT > 23$) in a large part of the diagram, below the white line in Figure 1e. In this region, no annihilation of skyrmion bubbles are observed from a few minutes and up to much longer timescales. The position $p_1$ near this white line has been selected for the observation of skyrmion bubbles under electric field.

The behavior of skyrmion bubbles under electric field gating is studied through a transparent $ITO$ electrode (see Figure1b). Polar Kerr images recorded for different applied electric fields are presented in Fig 2 a-c. A distribution of bubble diameters due to the presence of local inhomogeneities and pinning sites is observed at fixed electric field. The variation of skyrmion bubbles density with the electric field has been characterized by image processing (see Figure 2f). To explain the observed behavior, the analytical skyrmion bubble model described previously is used. The $t$, $M_s$ and $\sigma_w$ parameters and their variations with the electric field, used as an input for the simulation, are determined experimentally (see supplementary S3 and S8). When the voltage is switched from -5 V to +5 V, $M_s$ and $\sigma_w$ vary by respectively $+0.1$ MA/m and $+0.32$ mJ/m$^2$. These experimental parameters are used in the simulation in Figure 2g where the energy difference between a ferromagnetic state and an isolated bubble is plotted as a function of the bubble diameter. The observed decrease in the skyrmion bubbles number with increasing electric field is consistent with the increase in the nucleation energy barrier $E_n$ combined with a decrease of the annihilation energy barrier $E_a$ obtained in the simulation. In Figure 2 d,e differential images are shown where the blue/red bubbles correspond to bubbles which have appeared/disappeared. In Figure 2d, both blue and red bubbles are observed. This is explained by a domination of nucleation processes as $E_n \leq 23\ k_B T_{293K}$ and $E_a \geq 23\ k_B T_{293K}$ in the -5 V to 0 V range (Figure 2g). The observation of both blue and ref bubbles in that voltage range is a consequence of the displacement of existing bubbles due to a continuous nucleation of bubbles. On the contrary the $0\ V$ to $+5\ V$ range is dominated by annihilation processes with $E_a \leq 23\ k_B T_{293K}$ and $E_n \geq 23\ k_B T_{293K}$ as confirmed by a large majority of red bubbles in Figure 2e.

Now that the electric field control of skyrmion bubble density is demonstrated, we carry out the proof of concept of an electrical skyrmion switch by turning the electric field to higher voltages. In Figure 3a-b, we show the electric field switching from a state with a very high skyrmion bubbles density to a state with no skyrmion bubbles. This electric switching of skyrmion bubbles is reversible and reproducible sequentially as we can see in Figure 3d where the average Kerr intensity is recorded below the electrode as function of time while the voltage is switched between $+/-20$ V (see supplementary S9). To understand the origin of this switching effect, polar Kerr hysteresis loops with in plane applied magnetic field have been measured to estimate the variation of $M_S$ and $K_{eff}$ under electric field (see supplementary S8). When the voltage varies from $+20$ V to $-20$ V, $M_S$ and $K_{eff}$ are modified respectively by -47% and -64%. We deduce from $K_{eff} = K_s/t - 1/2\mu_0 M_S^2$ a $K_s$ variation $\Delta K_s/\Delta E = 830$ fJ/V/m. The electric field variations of the magnetic properties is most likely due to a modification of the electron density of state in the $Co$ and enhanced by the proximity of $T_C$ as observed in previous works on $Pt/Co/oxide$ in the ultrathin regime[37]. A variation of $D$ with electric field may also occur in our experiment as predicted theoretically[50]. It is possible to extract D from the expression $\sigma_W = 4\sqrt{A_{ex} K_{eff}} - \pi D$ using the experimental values of $K_{eff}$ and $\sigma_W$. However, as $M_s$ is varying with the electric field (due to the variation of $T_C$) the exchange constant $A_{ex}$ is also expected to vary and the detailed analysis necessary to separate the D variation from $A_{ex}$ is beyond the scope of this study. The variation of the skyrmion bubble energy with electric field (Figure 3c,d) is calculated using the experimental $M_s$ variations with electric field and a DW energy variation adjusted to fit the observed skyrmion bubble nucleation/annihilation rates (see supplementary S10). For positive voltage, skyrmion bubbles vanish due to a strong increase of the nucleation energy $E_n$ combined with a decrease of the annihilation barrier: nucleation is prevented and annihilation of bubbles is thermally activated. On the contrary, for a negative electric field, the nucleation energy $E_n$ is strongly reduced whereas $E_a$ becomes very large, thus favoring the creation of stable skyrmion bubbles. Consequently, the electric field can be used similarly to an applied magnetic field to switch skyrmion bubbles on and off by modifying the skyrmion stability. In our system the observed switch effect is associated to strong electric field variations of $M_s$ and $\sigma_W$ only expected near $T_C$. However, the variation of $M_s$ and $\sigma_W$ have opposite effects: an increase of $M_s/\sigma_W$ is decreasing/increasing $E_n$. Consequently, the skyrmion switch effect should be much more efficient at fixed $M_s$. To check this we calculated in Figure 3f the skyrmion bubble energy in the case of a thicker $Co$ layer of 0.6 nm, which is expected to show negligible $M_S$ variation with electric field (higher $T_C$) but a significant $K_{eff}$ change[51]. We see that the skyrmion switch effect (i.e. reducing the nucleation barrier and increasing the annihilation barrier) can be efficient at fixed $M_s$ with a $\sigma_W$ variation as small as 5%. In this case, the skyrmion bubbles are scaled down to a few 100 nm of diameter. Below this size, our thin wall analytical model is no more valid and further simulations are necessary to study the skyrmion swich effect on a nanometer sized skyrmion in a confined geometry.

In conclusion, we have reported the presence at RT of micron sized skyrmion bubbles in a $Pt/Co/oxide$ trilayer. The DMI strength, which is of the order of 1-2 mJ/m², reduces the DW energy by a factor of ~5 with respect to the classical Bloch case and enables the stabilization of skyrmion bubbles. The observed skyrmion bubbles density variation along the wedge and versus electric field is well described by an analytical isolated bubble model. We demonstrate the efficient and reproducible electric field writing and deleting of skyrmion bubbles. These results constitute a potentially important milestone towards the use of skyrmions for memory or logic devices.

**Methods**

**Sample preparation**

The $Pt(3 \text{ nm})/Co$ (0.6 nm)/$Al$ 0.76 nm to $Al$ 1.76 nm film was deposited by magnetron sputtering on a Si wafer with a ~500 nm thick thermal $SiO_2$ top layer. The $Al$ layer was oxidized in-situ by a $O_2$ plasma.

A ~90 nm $HfO_2$ layer is deposited ex-situ by Atomic Layer Deposition. Indium Tin oxide electrodes deposited by sputtering was patterned by lift-off in the form of $50 \times 800$ µm rectangles.

**Analytical model of the labyrinthine domains width**

The theoretical labyrinthine domain width $L$ has been calculated using a parallel band domains model in the case where the DW width is negligible compared to the domain diameter. For ultra-thin films, when the thickness is much smaller than the characteristic dipolar length[39,52] : $t \ll L_0$ with $L_0 = \sigma_W/\mu_0.M_s^2$ where $M_s$ is the saturation magnetization and $\sigma_W$ the domain wall energy, the domain width $L$ can be written as $L = \alpha.t.\exp(\pi L_0/t)$ where $t$ is the magnetic film thickness and $\alpha$ a constant $\alpha = \exp(\frac{\pi b}{2} + 1) \sim 0.955$ where $b$ is the numerical evaluation of a series[39]. Using this expression we obtain the domain wall energy $\sigma_W = \ln(L/(\alpha.t)).\mu_0.M_s^2.t/\pi$.

**Analytical "thin wall" bubble model**

We have used an analytical formula to approximate the energy of an individual bubble of radius $R$ relative to the saturated state[22,53]. The model used here is valid for $Q = K_u/K_d > 1$, where $K_u$ is the uniaxial (surface) out of plane anisotropy and $K_d$ the magnetic dipolar energy. In our case we have $Q = 0.83/0.54 = 1.53$. It was shown that this standard theory is well applicable for $Q > 1.5$. The energy of an individual bubble relative to the saturated state can be written as:

$$\Delta E_{sb} = \sigma_w.t.2.\pi.R + 2\mu_0 M_s.H.t.\pi.R^2 - \mu_0 M_s^2.\pi.t^3.I(d)$$

where $\sigma_w$ is the DW energy, $t$ the layer thickness, $R$ the bubble radius, $M_s$ the saturation magnetisation, $\mu_0 H$ the applied magnetic field, and $I(d)$ defined as follow:

$$I(d) = -\frac{2}{3\pi}d[d^2 + (1-d^2)E(u^2)/u - K(u^2)/u]$$

where $d = 2R/t$, $u^2 = d^2/(1+d^2)$ and $E(u)$ and $K(u)$ are elliptic integrals defined as:

$E(u) = \int_0^{\pi/2} \sqrt{1 - u.\sin^2(\alpha)} d\alpha$ and $K(u) = \int_0^{\pi/2} d\alpha/\sqrt{1 - u.\sin^2(\alpha)}$

**Acknowledgement**

This work was supported by the French National Research Agency (ANR) under the project ELECSPIN ANR-16-CE24-0018. We acknowledge the support from the Nanofab facility and the pôles magnetometrie and ingénierie experimentale from Institut Néel. We thank L. Cagnon for ALD deposition.

The authors thanks A. Thiaville, O. Fruchard, M. Chshiev, O. Boulle, I. M. Miron, N. Reyren, V. Cros, A. N. Bogdanov and B. A. Ivanov for fruitful discussions.

**Supporting information**

Sample characterisation. Current induced skyrmion bubbles motion. Real time room temperature spontaneous nucleation and annihilation of skyrmion bubbles. Sample characterisation under electric field. Switching of skyrmion bubbles with electric field. Estimation of the electric field variation of the domain wall energy using the isolated bubble model.

**Competing financial interests**

The authors declare no competing financial interests.

**Corresponding author**

Correspondence to: anne.bernand-mantel@neel.cnrs.fr

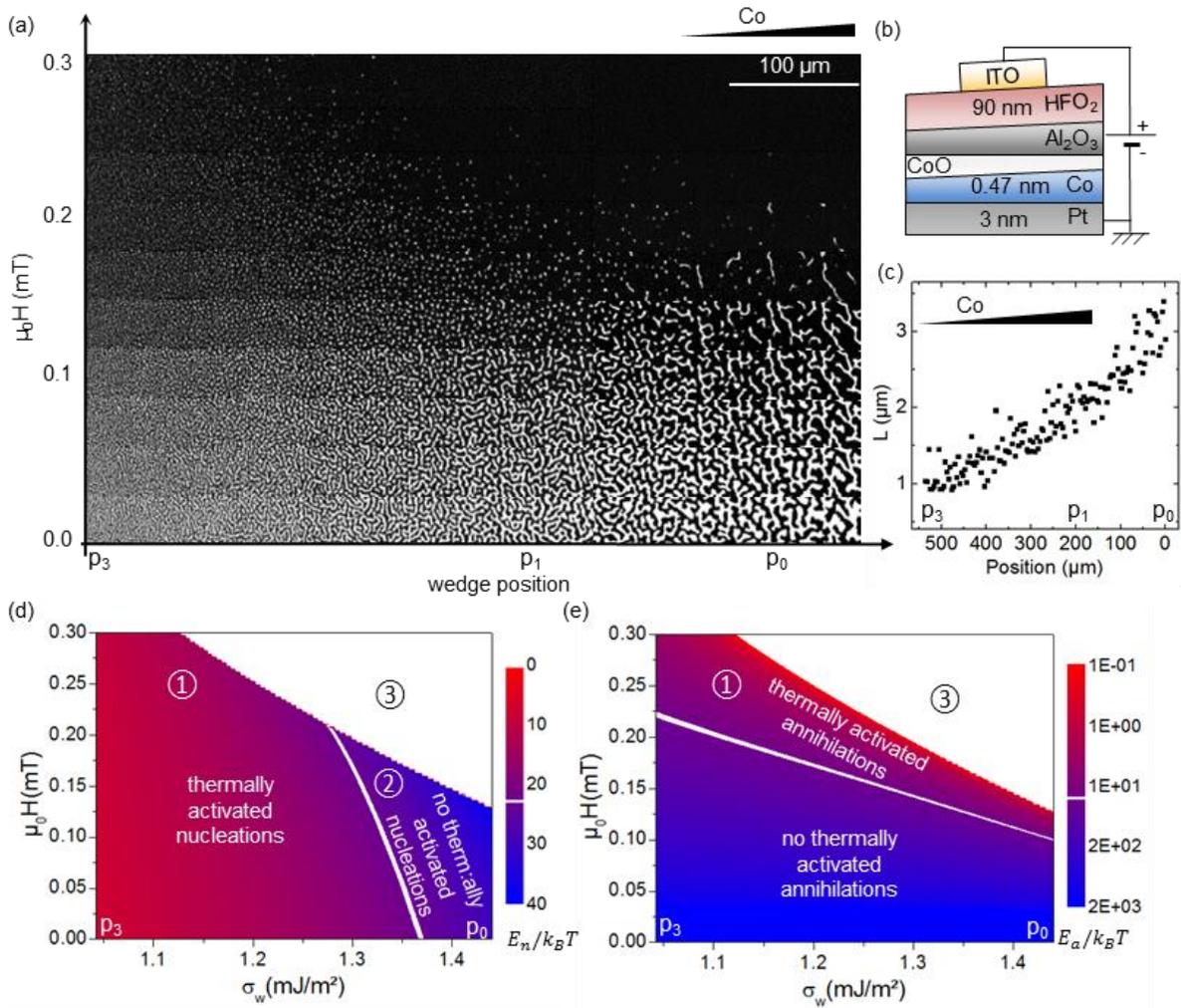

**Figure 1:** (a) Polar Kerr magnetic images of the labyrinthine and skyrmion bubble states in the Pt/Co/oxide trilayer. The image is made of 10 images of 30 x 550 µm. Each image is recorded a few seconds after applying an out of plane magnetic field. The magnetic field was varied from 0 to 0.3 mT. The Co thickness increases from left to right. (b) Schematic representation of the device: the Pt/Co/oxide trilayer is covered by a 90 nm $HfO_2$ layer and a top Indium Tin Oxide (ITO) electrode. The Co thicknesses variation is of the order of 1% over the 500 µm wide observed region. (c) Characteristic domain width $L$ as function of the wedge position for zero applied magnetic field extracted by fast Fourier transform on polar Kerr images. (d) and (e) Skyrmion bubble nucleation (d) and annihilation energies (e) calculated with the isolated bubble model as function of out of plane magnetic field and DW energy using the $H$, $t$, $M_s$ and $\sigma_w$ parameters estimated experimentally.

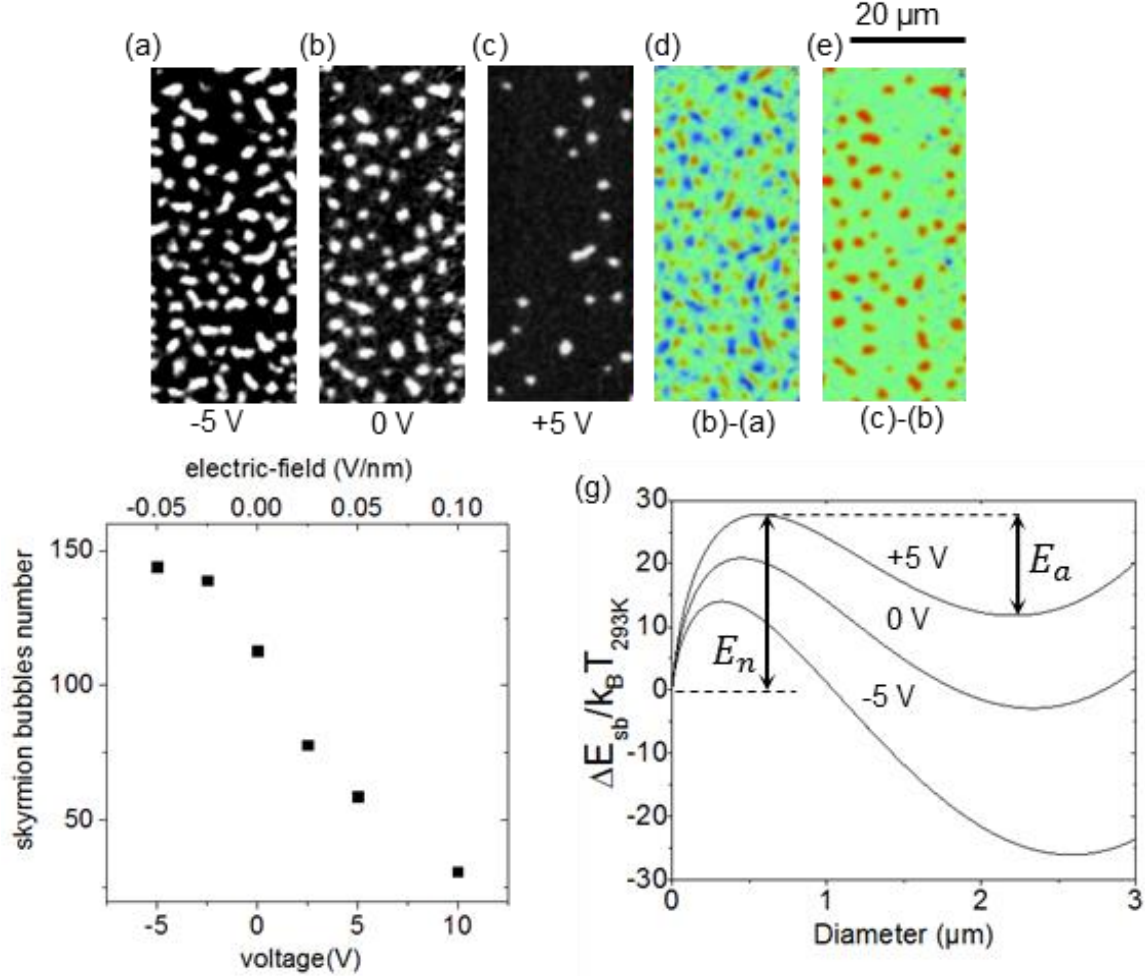

**Figure 2:** (a) to (c) Polar Kerr magnetic images of the electric field control of skyrmion bubbles density in the Pt/Co/oxide trilayer under a static $0.15\,\text{mT}$ perpendicular magnetic field, recorded through the transparent ITO electrode near position $p_1$ with different applied voltages. The sample is first saturated with a higher magnetic field, then the magnetic field is fixed to $0.15\,\text{mT}$ and the electric field is varied from $-5\,\text{V}$ to $+10\,\text{V}$. (d) and (e) Differential images obtained from (b)-(a)=(d) and (c)-(b)=(e). The blue/red bubbles correspond to objects which appeared/disappeared during the few seconds separating the images acquisitions. (f) Number of skyrmion bubbles extracted from images with twice the size of (a),(b) and (c). (g) Analytical calculation of the energy difference between a saturated state and a single isolated magnetic bubble as function of the bubble diameter with the parameters $t = 0.468\,\text{nm}$, $\mu_0 H = 0.15\,\text{mT}$, $M_s = 0.92\text{+/-}0.05\,\text{MA/m}$ and $\sigma_w = 1.33\text{+/-}0.16\,\text{mJ/m}^2$ corresponding to respectively $0\,\text{V}, +5\,\text{V}$ and $-5\,\text{V}$. The nucleation $E_n$ and annihilation $E_a$ energies are indiczted for the $+5\,\text{V}$ case. The $t, H, M_s$ and $\sigma_w$ parameters used for the simulation are determined experimentally (see supplementary S3, S8)

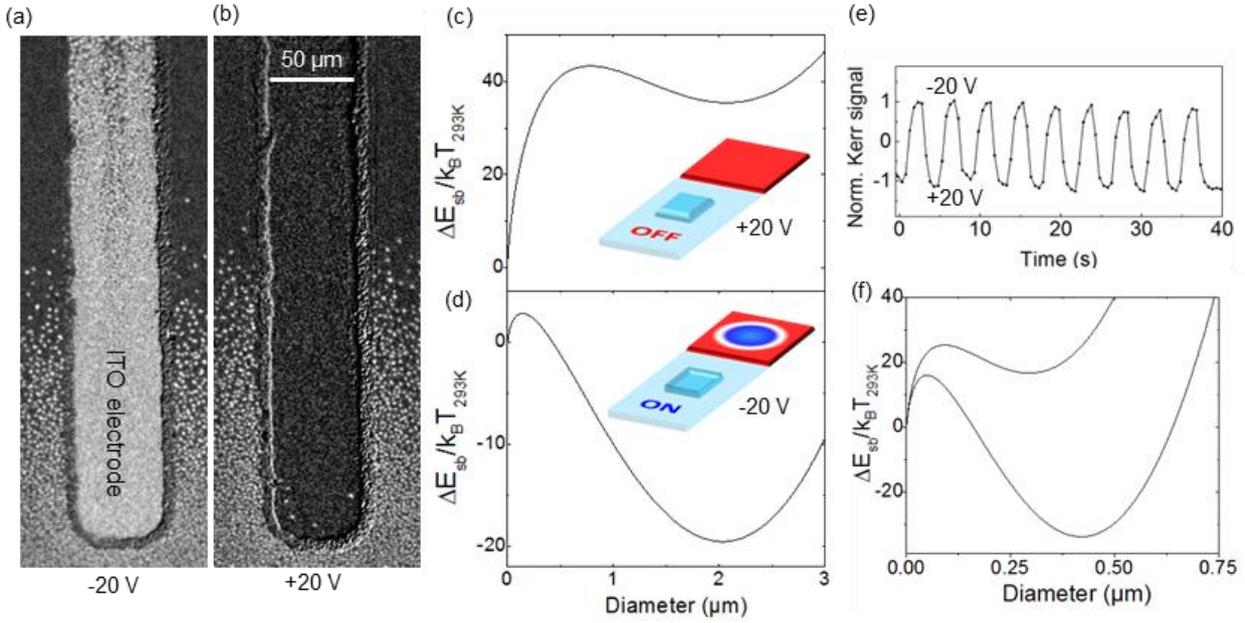

**Figure 3:** (a) and (b) Polar Kerr images of the electric field switching of skyrmion bubbles in the Pt/Co/oxide trilayer under a static 0.15 mT perpendicular magnetic field recorded 2 s after applying respectively -20 V and +20 V. (c) and (b) Simulation of the skyrmion switch effect using an analytical isolated skyrmion bubble model with $t = 0.468$ nm, $\mu_0 H = 0.15$ mT, $M_s = 1.07$ MA/m (+20V), $M_s = 0.57$ MA/m (-20V), $\sigma_w = 1.87$ mJ/m² (+20V) and $\sigma_w = 0.45$ mJ/m² (-20V). (e) Polar Kerr signal recorded through the electrode as a function of time during a sequence of electric field switching from +20 V to -20 V repeated 10 times every 2 s. (f) Simulation of the skyrmion switch effect in the case of a thicker Co layer with the parameters $t = 0.6$ nm, $\mu_0 H = 0.9$ mT, $M_s = 1.3$ MA/m, $\sigma_w = 2.72$ mJ/m² (top) and $\sigma_w = 2.57$ mJ/m² (bottom).

# The skyrmion switch: turning magnetic skyrmion bubbles on and off with an electric field

Marine Schott[1,2,3,4], Anne Bernand-Mantel[1,2]\*, Laurent Ranno[1,2], Stéfania Pizzini[1,2], Jan Vogel[1,2], Hélène Béa[1,3,4], Claire Baraduc[1,3,4], Stéphane Auffret[1,3,4], Gilles Gaudin[1,3,4] and Dominique Givord[1,2]

[1] Univ. Grenoble Alpes, F-38000 Grenoble, France

[2] CNRS, Inst. NEEL, F-38000 Grenoble, France

[3] CNRS, SPINTEC, F-38000 Grenoble, France

[4] CEA, INAC-SPINTEC, F-38000 Grenoble, France

\*anne.bernand-mantel@neel.cnrs.fr

**S1. VSM-SQUID magnetization measurements.**

The $Pt$ (3 nm)/$Co$ (0.6 nm)/$Al$ 0.76 nm to $Al$ 1.76 nm $+ O_2$ plasma sample has been cut to carry out VSM-SQUID measurements. The piece where the labyrinthine domains discussed in the paper appear is called sample A1 and has dimensions of $\sim 1$ mm $\times$ 4 mm. The A1 sample is represented schematically in Fig. S1 where the $p_0$, $p_1$, and $p_3$ positons are the one indicated also in Fig. 1a. The magnetic moment is measured with a Quantum Design MPMS 3 VSM-SQUID equipment. The sample A1 was measured 4 times at room temperature and we find $M_s.t = 4.2 \pm 0.4 \times 10^{-4}$ A/m.m. As a part of the $Co$ has been oxidized by $O_2$ plasma in the sample A1, the $Co$ thickness is lower than the initial thickness $t = 0.6$ nm. The temperature dependent magnetization measurements, carried out in the 100-360K temperature range, on a sample similar to A1, are presented in Fig. S2. Each point correspond to the remanent magnetisation extracted from a M(H) loop. A Curie temperature $T_C = 325$ K is found using a combination of two power laws : $M(T/Tc)/M(0) = [1 - s(T/Tc)^{(3/2)} - (1-s)(T/Tc)^{(5/2)}]^{1/3}$ with s=0.11 for Co[1]. An $M_s$ decrease of 38% between 0 K and 293 K is observed. This magnetization variation is used to define the lower bound for the Co thickness: we assume that the $M_s(0 \text{ K})$ cannot exceed the bulk Co value of $1.44 \times 10^6$ A/m which gives $M_s(293\text{K}) < 0.89 \times 10^6$ A/m. We deduce a lower bound for the thickness $t > 0.47$ nm from the $M_s.t$ value.

**S2. Polar Kerr measurements along sample A1.**

To estimate the $M_s$ variation along the sample A1, local Kerr measurements were carried out every 160 µm. In Fig. S3b, we show the polar Kerr measurements with in plane applied magnetic field. This measurement correspond to the out of plane projection of the magnetization while it rotates reversibly outside its easy axis toward its hard axis direction (see Fig. S5a). We define the maximum Kerr signal amplitude as the difference between the maximum and the minimum intensities. This Kerr signal amplitude is proportional to the Kerr rotation, itself proportional to $M_s.t_{Co}$. The Kerr signal maximum amplitude is plotted in Fig. S3a as function of the wedge position. As the change in the Kerr signal intensity (<1%) and the change of the Co thickness (1%) are negligible between positions 1 and 5, we assume that the 17% drop in the Kerr signal is related to a $M_s$ variation. A small change of the Co thickness can lead to a significant change in $M_s$ due to the proximity of the $T_C$. A similar effect was observed in $Pt/Co/Al_2O_3$ in the work by Chiba et al.[2] where they observe a change in the $T_C$ by few tens of K when the Pt thickness is modified by few %.

**S3. Determination of $M_s$, $t$ and $\sigma_w$ and their variation along the wedge.**

As the $M_s$ and $t$ parameters are known with a significant uncertainty (of the order of +/−10%) we have checked that a variation of these parameters, between $M_s = 0.85\ to\ 1.05$ A/M and $t = 0.43\ to\ 0.51$ nm does not change any of the conclusions of the paper. In particular, only minor differences in the simulations of Fig. 1d,e, Fig. 2g, Fig. 3c,d are found. Consequently, we decided to fix the average value of $M_s(293K)$ for sample A1 to its upper bound $0.89 \times 10^6$ A/m obtained in paragraph S2. The variation of $M_s$ along the sample A1 is obtained by a fit to the Kerr data in Fig. S3a using $M_s = 0.93 \times 10^6 \text{-} 0.27x^2$ where $x$ is the position in µm. The corresponding $M_s$ values are shown in the right y axis of Fig. S3a. The thickness is fixed also to its lower bound $t = 0.47$ nm. To estimate the variation of the Co thickness along the wedge we made the assumption that the Co thickness variation due to oxidation is a linear projection of the aluminium thickness variation (see Fig S1). The deposition is done by sputtering in a system with square targets. The wedge is obtained by progressively moving the shutter during the deposition. To calibrate the thickness variation over the 10 cm width of the substrate, a thicker wedged layer has been deposited and the thickness has been measured regularly using X-ray reflectivity along the sample. Assuming that the slope varies proportionally with the thickness, we deduce a slope of 0.1 nm/cm for the aluminium wedge, and consequently the Co layer in the overoxidized zone, assuming that the oxygen plasma is homogeneous.

In Fig. S4, we show the domain wall energy and its variations, calculated from the analytical expression of the labyrinthine domains width defined in the Methods : $\sigma_w = \ln(L/(0.955.t)).\mu_0 M_s^2.t/\pi$ using the $M_s$ and $t$ values given previously and the experimental domain width $L$ values extracted by fast Fourier transform from Fig.1a which is plotted in Fig.1c. These values are fitted with a polynomial expression: $\sigma_w = 1.44.10^{-3}\text{-}4.53.10^{-7}x\text{-}5.43.10^{-10}x^2$ where $x$ is the position in µm (Fig. S4). The table T1 summarizes the experimental $M_s$, $t$ and $\sigma_w$ values defined here and their variations along the wedge. These values will be used as an input for the isolated skyrmion bubble simulations in Fig. 1 d and e.

| | |
|---|---|
| $t$ (nm) | $0.47 \times 10^{-9}\text{-}0.01 \times 10^{-12}x$ |
| $M_s$ (MA/m) | $0.93 \times 10^6\text{-}0.27x^2$ |
| $\sigma_w$ (mJ/m$^2$) | $1.44 \times 10^{-3}\text{-}4.53 \times 10^{-7}x\text{-}5.43 \times 10^{-10} x^2$ |

**Table T1** : experimental $t$, $M_s$ and $\sigma_w$ values and their variations. $x$ is the position in µm with respect to position $p_0$.

## S4. Determination of $K_{eff}$ and $M_s$ variations using Kerr measurements combined with a Stoner-Wohlfarth model.

We carry out simulations of the $M(H)$ loops in the hard axis direction based on the Stoner-Wohlfarth (SW) model which allows to reproduce the magnetization rotation and estimate $K_{eff}/M_s$. We assume that the system has a uniaxial easy axis perpendicular to the film surface (Fig. S5). The energy per unit volume is: : $E = K_{eff}sin^2(\theta) - \mu_0 H M_s cos(\theta - \varphi)$ where $K_{eff}$ is the uniaxial anisotropy in $J/m^3$, $\theta$ the angle between the magnetization and the easy axis and $\varphi$ the angle between the magnetic field and the easy axis (see Fig. S5). An angle $\alpha$ is introduced to take into account the misalignment between the applied magnetic field, the film plane and the direction perpendicular to the observation direction (Fig. S5b). In Fig. S3b the normalised polar Kerr signals for in plane applied magnetic field measured at positions 1 and 3 are superimposed with the SW fits obtained with the parameters $M_s = 0.92$ MA/m (1), $M_s = 0.85$ MA/m (3), $K_{eff} = 0.28$ MJ/m$^3$ (1), $K_{eff} = 0.24$ MJ/m$^3$ (3), $\varphi = 84°$ and $\alpha = 2°$. At position 0 we fond $M_s = 0.93$ MA/m, $K_{eff} = 0.30$ MJ/m$^3$, $\varphi = 84°$ and $\alpha = 2°$.

## S5. Real time room temperature spontaneous demagnetization at zero applied magnetic field.

The video **V1** displays in real time the polar Kerr images of the room temperature spontaneous demagnetization. The sample was saturated with an out of plane magnetic field which is turned to zero just after the beginning of the video. We observe a spontaneous demagnetization of the sample in few seconds via the nucleation and growth of magnetic domains. When the sample is demagnetized, the thermally activated movement of the DW is still visible. Such thermal melting of stripe domains has been observed in systems near the spin reorientation transition[3–9].

## S6. Current induced skyrmion bubbles motion.

The sample was etched to form a Hall cross in order to check the unidirectional skyrmion bubbles displacement with electrical current. The channels of the Hall cross are 10 µm large. A DC current of 0.15 mA is applied between the top right and bottom contact. The video **V2** displays in real time the polar Kerr image of the Hall cross at room temperature under a ~0.15 mT constant magnetic field applied in the black direction. The skyrmion bubbles have a steady motion against the electron flow for a current density of $5 \times 10^9 \ A/m^2$. Some bubbles get pinned and/or distort into elongated domains reminiscent of stripe domains. This result is similar to what has been obtained in other systems presenting skyrmion bubbles[10–12]. A more detailed bubble motion study was not carried out in our work.

## S7. Real time room temperature spontaneous nucleation and annihilation of skyrmion bubbles.

In video **V3**, the room temperature real time polar Kerr image near the number 1 indicated in Fig. 1d-e under a constant out of plane magnetic field of ~0.25 mT applied in opposite direction to the bubbles magnetization. Spontaneous nucleation and annihilation of white skyrmion bubbles with a rate > 1 Hz are observed.

## S8. Experimental determination of the electric field variation of $M_s$, $K_{eff}$ and $\sigma_W$ using Kerr measurements and the Stoner-Wohlfarth model.

Polar Kerr loops with in plane applied magnetic field have been recorded at position $p_1 = 215$ µm under electric field to characterize the variations of $M_s$ and $K_{eff}$. The $M_s$ value at position $p_1$, determined in paragraph S3 is $M_s = 0.92$ MA/m. We use the SW model described in the paragraph S4 to deduce $K_{eff}$ and extract the $M_s$ and $K_{eff}$ variations with electric field. In Fig. S6b, the normalised polar Kerr signal for in plane applied magnetic field measured at positions $p_1$ with 0 V and -10 V applied voltages are superimposed with the SW fits obtained with the parameters $M_s = 0.92$ MA/m (0 V), $M_s = 0.82$ MA/m (−10 V), $K_{eff} = 0.27$ MJ/m³ (0 V), $K_{eff} = 0.22$ MJ/m³ (−10 V), $\varphi = 84°$ and, $\alpha = 5°$. In Fig.

S6a, a polar Kerr image is showing the effect of the electric field on the labyrinthine domains: the domains size is shifted by $\sim 400\ \mu m$ toward the thinner wedge direction for negative voltage (electron depletion in Co). This observation is consistent with the equivalence between a reduction of the electron density and an oxidation increase observed in our previous work[13]. The labyrinthine domain width $L$ and its electric field variation is extracted from the image of Fig. S6a using fast Fourier transform. Then, as in paragraph S3, we use this values to calculate the domain wall energy and its variation under electric field using the experimental $M_s = 0.92\ \text{MA/m}$ and $t = 0.468\ \text{nm}$ values and their variations with the wedge position, and $\Delta M_s = M_s(-10\ \text{V})-M_s(0\ \text{V}) = -0.1\ \text{MA/m}$ obtained from the SW fit in Fig. S6b. The obtained DW energy is plotted in Fig. S6c. A linear fit allows to extract the electric field variation of the DW energy $\Delta\sigma_w = \sigma_w(-10\ \text{V})-\sigma_w(0\ \text{V}) = -0.32\ \text{mJ/m}^2$.

In Fig. S7 the normalised polar Kerr signals for in plane applied magnetic field measured another day at positions $p_1$ for higher applied voltages +20 V and -20 V are superimposed with the SW fits obtained with the parameters $M_s = 1.07\ \text{MA/m}\ (+20\ \text{V}), M_s = 0.57\ \text{MA/m}\ (-20\ \text{V}), K_{eff} = 0.42\ \text{MJ/m}^3\ (+20\ \text{V}), K_{eff} = 0.15\ \text{MJ/m}^3\ (-20\ \text{V}),\ \varphi = 86°$ and $\alpha = 6°$.

## S9. Switching skyrmion bubbles with electric field.

The video **V4** displays, in real time, at room temperature, the polar Kerr image of the skyrmion switch effect at position $p_1$. The sample was saturated under a high out of plane magnetic field in the opposite direction (black) to the bubbles magnetization before the beginning of the video, then a $0.15\ \text{mT}$ out of plane magnetic field is applied in the black direction and kept constant. During the video the electric field is switched every two seconds between -20 V and + 20 V starting with + 20 V.

## S10. Estimation of the electric field variation of the DW energy $\sigma_w$ for high electric field using the isolated bubble model.

At high electric fields the labyrinthine domains disappear as the thermal activation is either too high (at -20 V), leading to a complete melting of the labyrinthine domains, or too low (at +20 V), preventing the system to demagnetize and the labyrinthine domains to form. Consequently, the experimental determination of the DW energy $\sigma_w$ using the labyrinthine domain width $L$ was not possible for high electric fields. To estimate $\sigma_w$ for these high voltages, we applied the isolated bubble model, using as an input the $M_s$ values for high electric field obtained in paragraph S8 from Kerr measurements under electric field. We start by the +20 V case. At this voltage all the skyrmion bubbles disappear, as observed in Fig. 3b and video V4. This means that the annihilation energy barrier (defined in Fig. 2g) is smaller than the RT thermal energy limit $E_a/k_B T_{293K} < 23$, and much smaller than the nucleation energy $E_a \ll E_n$. This is true for DW energies $\sigma_w > 1.85$ mJ/m² as shown in Fig. S8a. This gives a lower bound for the increase of $\sigma_w$ under electric field $\Delta\sigma_w^{pos}=\sigma_w(20\ V)- \sigma_w(0\ V)=1.85-1.33>0.52$ mJ/m². For -20 V, we observe a fast nucleation of skyrmion bubbles in Fig. 3a and video V4 and we deduce a nucleation energy $E_n/k_B T_{293K} < 23$, an annihilation energy $E_a/k_B T_{293K} > 23$. This gives a DW energy $\sigma_w < 0.47$ mJ/m² as shown in Fig. S8b and a lower bound for the decrease of $\sigma_w$ under electric field $\Delta\sigma_w^{neg} = \sigma_w(0\ V)- \sigma_w(-20\ V)=1.33-0.47>0.86$ mJ/m².

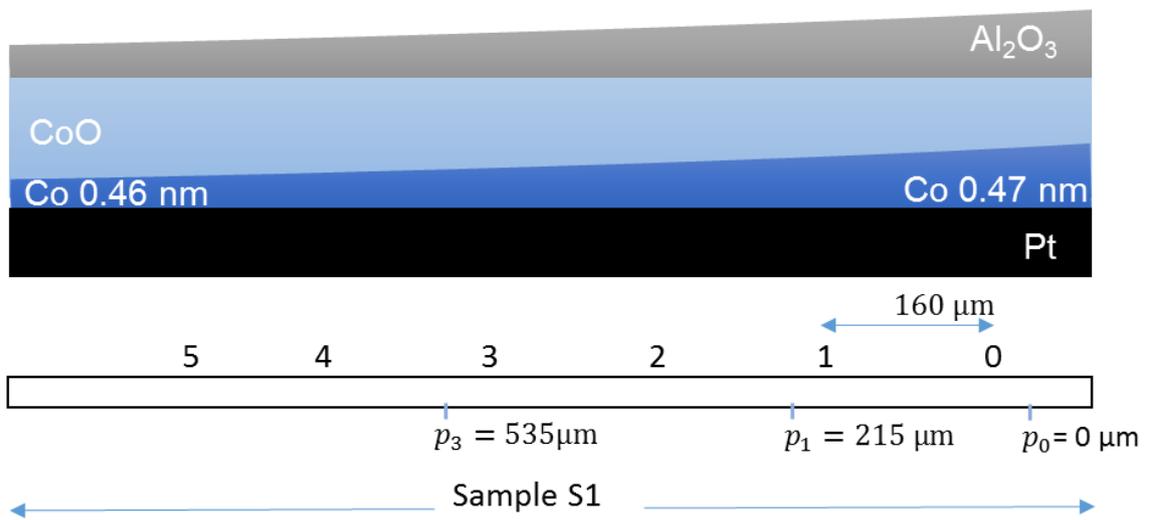

**Figure S1 :** Schematic representation of the sample A1 with the 0-5 and $p_0$-$p_3$ positions used in the text.

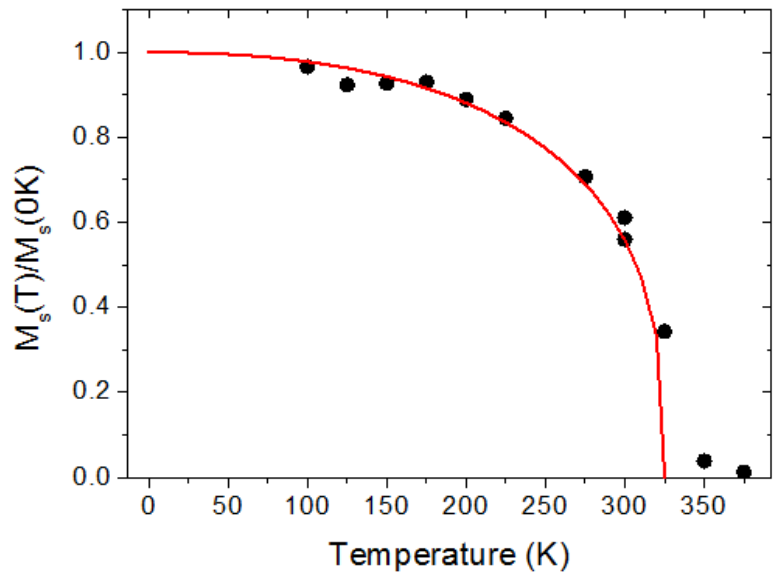

**Figure S2 :** VSM-SQUID measurements of the remanent magnetization of a sample similar to A1 as function of the temperature. Each point is the remanent magnetization extracted from M(H) loops.

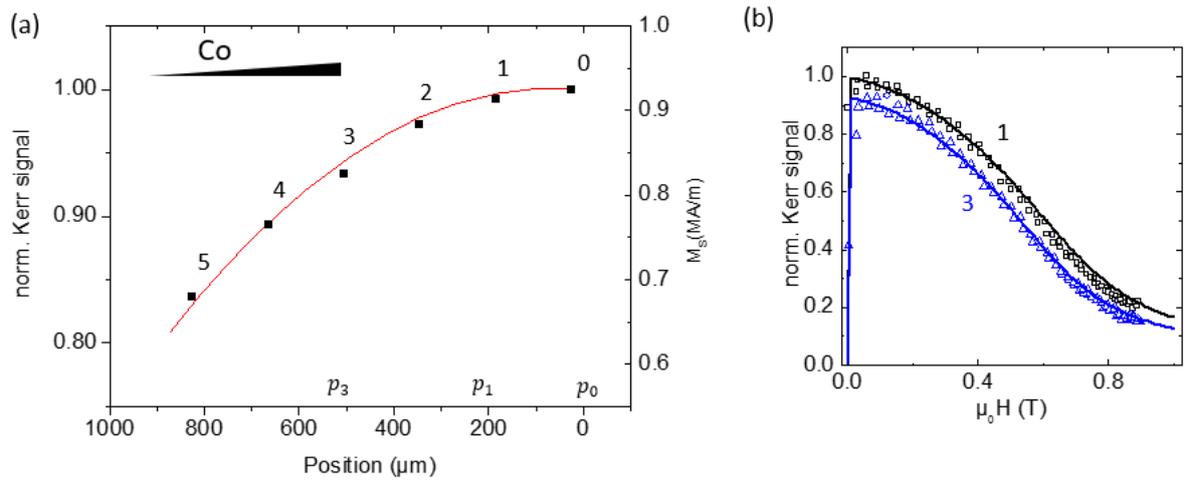

**Figure S3 : a**, Polar Kerr loop maximum amplitude recorded every 160 μm along the wedge starting from a position around $p_0$ divided by the Kerr loop maximum amplitude at position $p_0$. The corresponding magnetization values are indicated on the right y axis. **b**, Polar Kerr loops versus in-plane magnetic field recorded at positions 1 (black squares) and 3 (blue triangles) indicated on **a**. SW fit (black and blue lines, see S4).

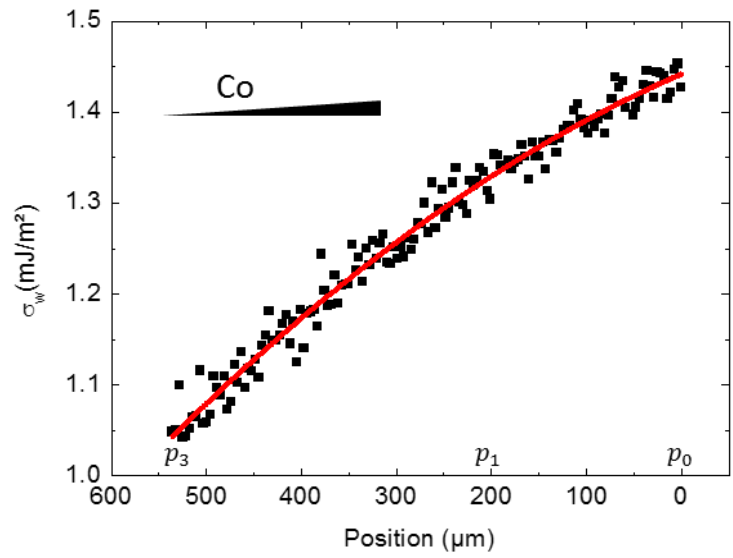

**Figure S4 :** Experimental values of the DW energy along the wedge calculated using the experimental $M_s$, $t$ and $L$ values (black squares) and polynomial fit (red line) (see S3).

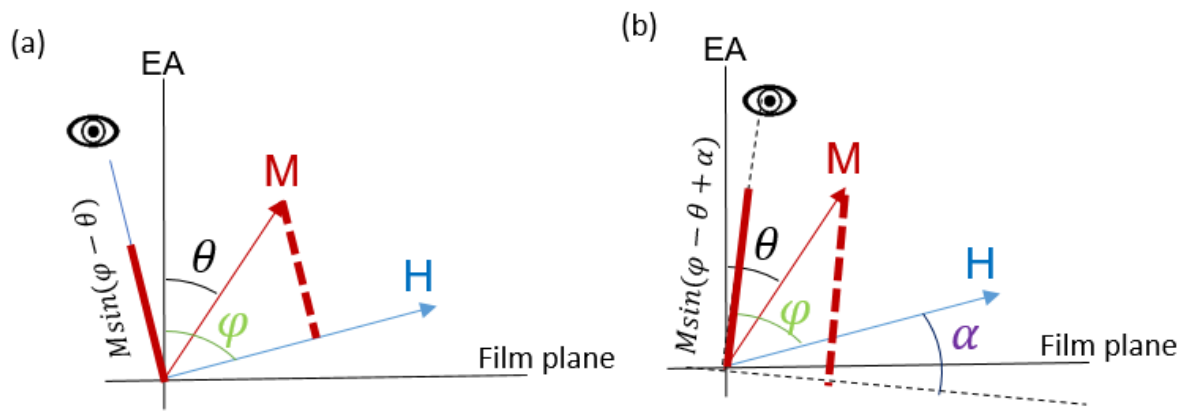

**Figure S5:** Definition of the angles between the magnetization, the applied magnetic field, the film plane and the observation direction (represented by an eye).

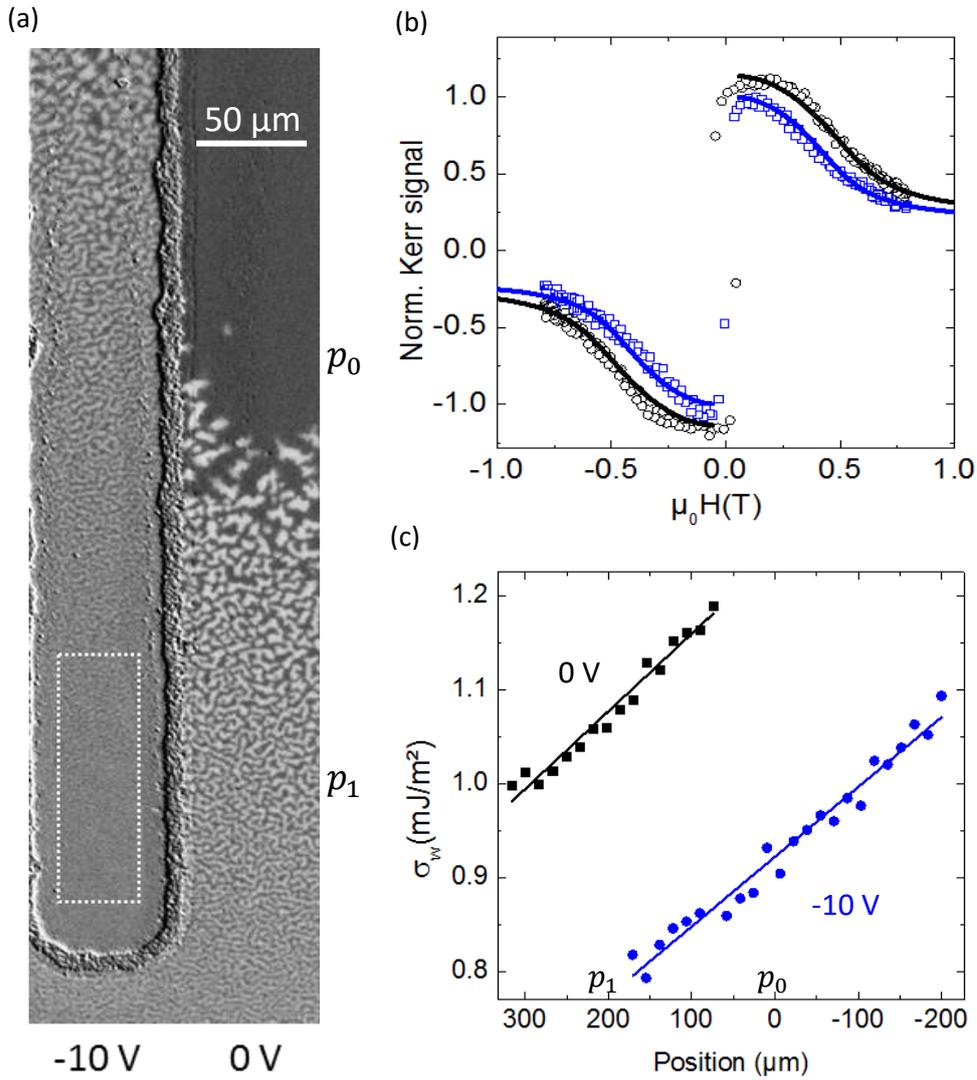

**Figure S6**: (a) Polar Kerr image of the labyrinthine domains under electric field at zero applied magnetic field. The dashed rectangle correspond to the zone where the Kerr signal of (b) was recorded **b**, Normalized polar Kerr signal for in plane applied magnetic field at positions $p_1$ on the wedge under applied voltages of 0 V (black circles) and -10 V (blue squares) superimposed with the SW fit (blue and black lines). **c**, Experimental (squares and dots) DW energies as function of the wedge position and for two applied voltages calculated from the analytical expression of the labyrinthine domain width using the $L$ values extracted by Fast Fourier transform from **a**. The black and blue lines are linear fits.

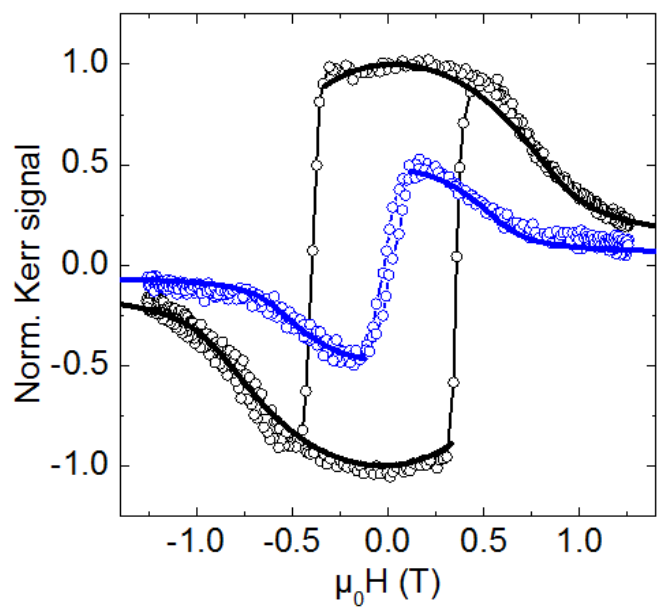

**Figure S7**: Normalized polar Kerr signal for in plane applied magnetic field at positions $p_1$ under applied voltages of $+20$ V (black circles) and $-20$ V (blue circles) superimposed with the SW fits (black and blue lines).

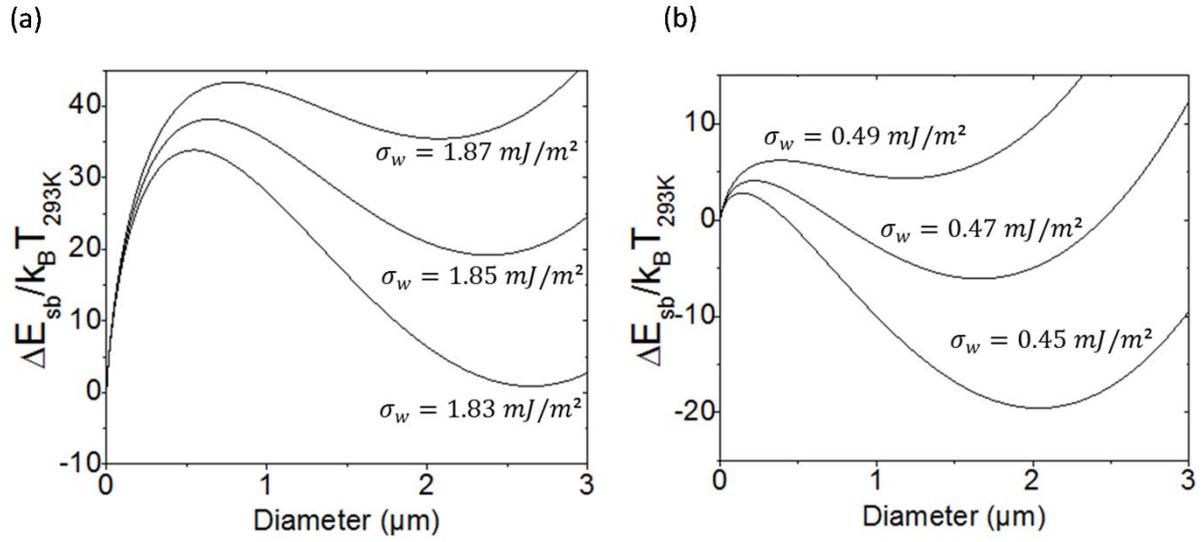

**Figure S8**: Isolated skyrmion bubble energy as function of the bubble diameter calculated with $t = 0.468$ nm, $\mu_0 H = 0.15$ mT and the magnetization values corresponding to (a) the +20 V case with $M_s = 1.07$ MA/m and (b) the -20 V case with $M_s = 0.57$ MA/m. The DW energy parameters are indicated in the graphs.